\documentclass[11pt,twoside]{article}
\usepackage{./asp2014}

\aspSuppressVolSlug
\resetcounters

\begin{document}

\title{The X-ray Background Emission of the Galactic Center and Bulge with {\it NuSTAR}}
\author{Ekaterina~Kuznetsova,$^1$ Roman~Krivonos,$^1$ Kerstin Perez$^{2,*}$ and Daniel~R.~Wik$^3$
\affil{$^1$Space Research Institute of the Russian Academy of Sciences (IKI), Moscow, Russia;}
\affil{$^2$Department of Physics, Massachusetts Institute of Technology, Cambridge, USA; \email{kmperez@mit.edu}}
\affil{$^3$Department of Physics and Astronomy, University of Utah, Salt Lake City, USA;}
\affil{$^{*}$Corresponding author}}

\paperauthor{Ekaterina~Kuznetsova}{eakuznetsova@cosmos.ru}{0000-0003-0938-5317}{}{Space Research Institute of the Russian Academy of Sciences (IKI)}{Moscow}{}{117997}{Russia}
\paperauthor{Roman~Krivonos}{krivonos@iki.rssi.ru}{0000-0003-2737-5673}{}{Space Research Institute of the Russian Academy of Sciences (IKI)}{}{Moscow}{117997}{Russia}
\paperauthor{Kerstin~Perez}{kmperez@mit.edu}{0000-0002-6404-4737}{Massachusetts Institute of Technology}{Department of Physics}{Cambridge}{MA}{02139}{USA}
\paperauthor{Daniel~R.~Wik}{wik@astro.utah.edu}{0000-0001-9110-2245}{University of Utah}{Department of Physics and Astronomy}{Salt Lake City}{UT}{84112}{USA}

\begin{abstract}
The Galactic diffuse X-ray  emission (GDXE) is believed to arise from unresolved populations of numerous  low-luminosity X-ray binary systems that trace stellar mass distribution of the Milky Way. Many  dedicated  studies  carried out over the last decade suggest that a dominant contributor to GDXE is a population of accreting white dwarfs (WDs). The question arises about relative contribution of different subclasses of accreting WD population, namely non-magnetic WD binaries, magnetic intermediate polars (IPs) and polars, in different regions of the Galaxy: the Galactic center, bulge, and ridge. Recent low-energy (E$<10$~keV) studies indicate that non-magnetic WD binaries, in particular quiescent dwarf novae, provide a major contribution to the diffuse hard X-ray emission of the Galactic bulge. From the other side, previous high energy (E$>10$~keV) X-ray measurements of the bulge and ridge imply a dominant population of magnetic CVs, in particular intermediate polars. In this work we use side aperture of the {\it NuSTAR} to probe the diffuse continuum of the inner $\sim1-3^{\circ}$ of the Galactic bulge, which allows us to constrain possible
mixture of soft and hard populations components of the spectrum. We found that GDXE spectrum is well-described by a single-temperature thermal plasma with $kT \approx 8$~keV, which supports that the bulge is dominated by quiescent dwarf novae with no evidence of a significant intermediate polar population in the hard X-ray band. We also compare this result with previous {\it NuSTAR} measurements of the inner 10~pc and inner 100~pc of the Galactic center.
\end{abstract}

\section{Galactic bulge diffuse X-ray emission with {\it NuSTAR}}

The Galactic diffuse X-ray emission (GDXE) was discovered more than 30 years ago. It extends along the Galactic plane over $100^{\circ}$ and fills the Galactic center. It is convenient to consider following distinct regions of the GDXE: the Galactic center (GCXE) in the inner $\sim$100~pc or $|l|\lesssim0.^{\circ}5$, the Galactic bulge (GBXE) in the inner $\sim$1~kpc or $|l|\lesssim5^{\circ}$, and the Galactic ridge (GRXE) $|l|\approx5^{\circ}-100^{\circ}$. 

Origin of the GDXE emission is still not fully clear. Soon after discovery, the main question was whether GDXE is truly diffuse or consists of a large number of unresolved sources. The broad-band spectral and large-scale morphology analysis of the GDXE revealed that its nature is mainly due to low-luminosity unresolved point sources \citep{revnivtsev06, revnivtsev09}. A truly diffuse origin was found unrealistic, because plasma must have too high temperature $\sim8~keV$ to be held in Galaxy's gravitational potential \citep[see, e.g.][]{tanaka02,ebisawa05}. 

The studies of the GDXE in X-ray band are complicated due to several reasons. The instrument must have  large field of view (FOV) to collect weak GDXE from the large regions of the Galaxy. The second reason is a difficulty of the separation of photons detected from different parts of the GDXE. For example, {\it RXTE} and {\it INTEGRAL} can provide large enough FOV but their spatial resolution is not sufficient for detail analysis of these regions. 

{\it NuSTAR} is a focusing telescope, which operates in the hard X-ray energy band 3--79~keV. {\it NuSTAR} FOV is about 13'' for focusing optics, but it also allows to detect unfocused photons from a large side aperture \citep{perez17}. It provides a unique ability to study the innermost Galactic center emission and the diffuse emission of the bulge with the same instrument.

Observations of the Galactic center region with {\it NuSTAR} used in this work were carried out during the period from July 2012 through October 2014 with a total exposure $\sim$2~Ms \citep[see, e.g.][]{mori15}. \cite{perez19} obtained GDXE spectrum from the {\it NuSTAR} FMPA focal plane detector not contaminated by stray-light from nearby bright X-ray sources. The collected spectrum was constructed from focused and unfocused aperture photons and instrumental background. Spectral analysis required a complicated modeling of all spectral components, including GBXE, GCXE, cosmic X-ray background (CXB) and instrumental background \citep[see details in][]{perez19}.

The unfocused GBXE component was investigated with a single-temperature plasma (1T) model and an intermediate polars mass (IPM) model (Fig.~\ref{fig:GB}). This approach allows to directly compare {\it NuSTAR} measurements with a previous {\it NuSTAR} observations of the GC and {\it Suzaku}, {\it RXTE} and {\it INTEGRAL} investigations of the GBXE and GRXE. The best-fit plasma temperature $kT$ and average white dwarf mass $M_{\rm WD}$ was estimated at $kT\approx8~keV$ and $M_{\rm WD}=0.4-0.5M_{\odot}$ for the 1T and IPM models, respectively. The spectrum of the GBXE was found significantly softer than the Galactic center spectrum  with an average temperature $kT>20~keV$ obtained with {\it NuSTAR} \citep{hailey16, perez15} and the previously observed {\it Suzaku} broad-band spectrum \citep{yuasa12}. However, from the other side, \cite{perez19} result is well consistent with {\it Suzaku} measurements at low-energies \citep[see, e.g.][]{yamauchi16, nobukawa16, koyama18}. This soft spectrum indicates that the main source population of the Galactic bulge is probably dwarf novae, rather than intermediate polars.

\articlefiguretwo{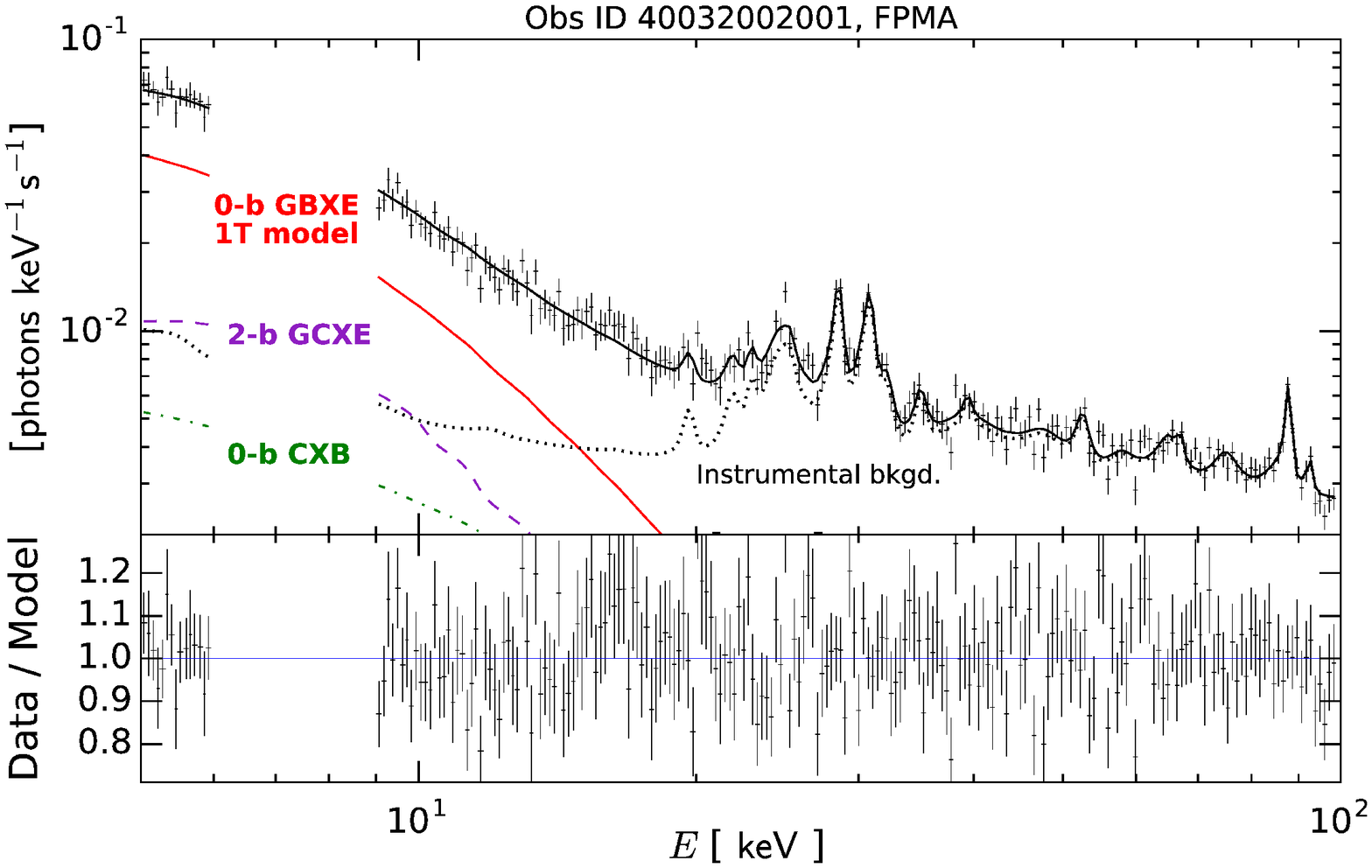}{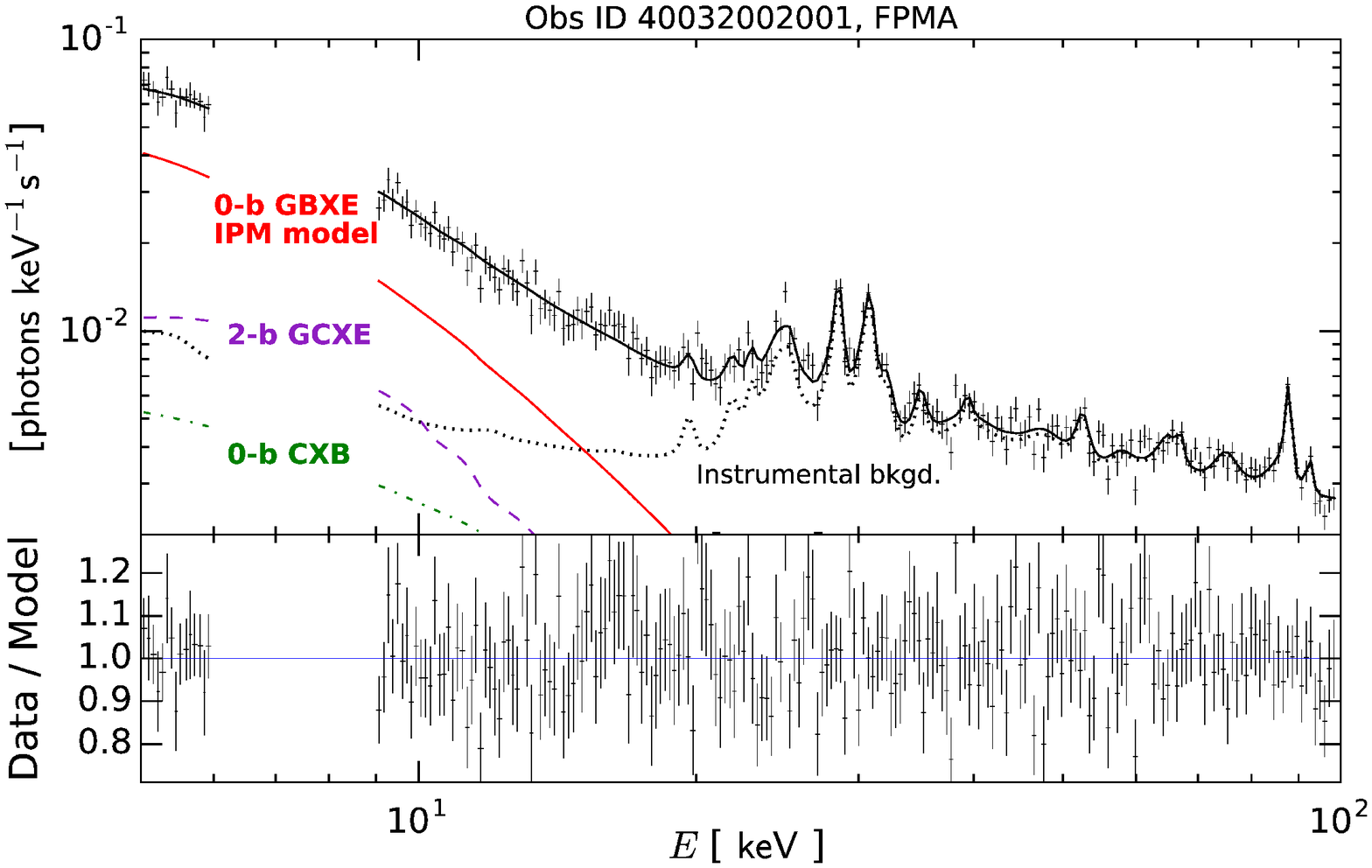}{fig:GB}{\emph{Left:} Data and folded best-fit model spectral components (solid unfocused GBXE, dashed focused GCXE, dash-dot unfocused CXB, focused CXB below y-axis range, dotted detector background) for FPMA of 400032002001 {\it NuSTAR} observation. The unfocused GBXE component is described by the 1T model. \emph{Right:} The same as left, but the unfocused GBXE component described by the IPM model. Adopted from \cite{perez19}.}

\section{Discussion and comparison with GCDE}
\label{sec:GC}

\cite{perez15} discovered a faint diffuse extended emission in the inner part of the GC at energies above 20~keV with {\it NuSTAR} optics. This emission, referred as GCDE, was detected up to 40~keV with its peak located at the position of Sgr A* and extended along the Galactic plane (Fig.~\ref{fig:GC}). The 2D image analysis of GCDE morphology showed that the hard X-ray emission is significantly narrower in both longitude and latitude than the soft X-ray distribution studied by \cite{heard13} with {\it XMM-Newton}. 

The hard emission component of the GCDE above 20~keV is well described by either a power-law model with the photon index of $\sim1.2-1.9$ or a poorly constrained high-temperature bremsstrahlung model with the plasma temperature $kT>35~keV$.  \cite{perez15} concluded that GCDE spectral properties are consistent with a population of intermediate polars with a mean white dwarf mass $M_{\rm WD}>0.9M_{\odot}$. This population is more massive than previously observed in the Galactic center and ridge \citep[$M_{\rm WD}\approx0.5M_{\odot}$][]{muno04,krivonos07,heard13} and in the Galactic bulge \citep[$M_{\rm WD}\approx0.66M_{\odot}$][]{yuasa12}.

\articlefigure[width=0.8\textwidth]{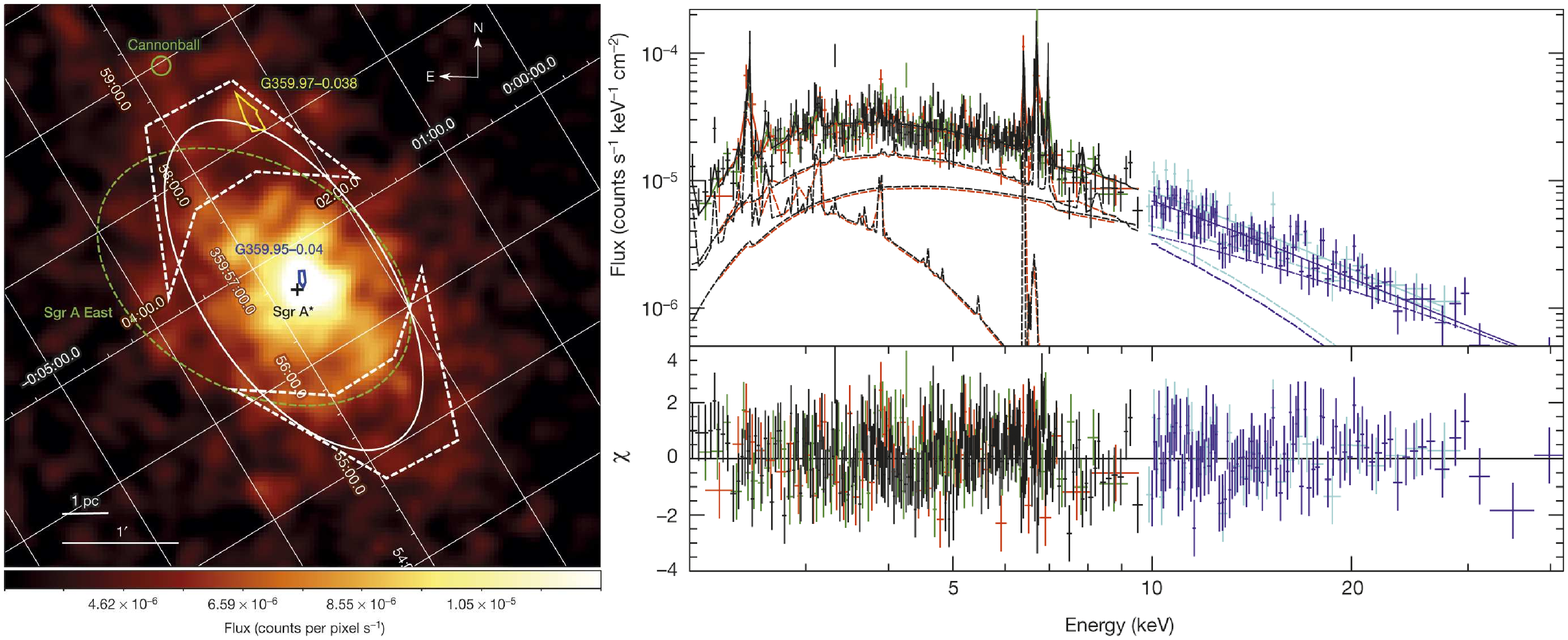}{fig:GC}{ \emph{Left:} The image in the 20--40~keV energy band of the inner $12~pc\times12~pc$ ($5'\times5'$) of the Galaxy. Solid ellipse demonstrates the FWHM of the best-fit 2D Gaussian. Energy spectra were extracted from dashed polygons. \emph{Right:} The 2--10~keV {\it XMM-Newton} spectrum and 10--40~keV {\it NuSTAR} spectrum. Dashed lines demonstrate different model components. Adopted from \cite{perez15}.}

The derived $kT \approx 8$~keV temperature of the GBXE with {\it NuSTAR}, is significantly lower than observed in the inner 10~pc and inner 100~pc of the Galactic center. It points out to that the diffuse hard X-ray emission of the Galactic center is dominated by IPs and emission of the Galactic bulge is dominated by DNe. The last is also consistent with recent {\it Suzaku} studies of the Fe line properties and low-energy continuum of the bulge \citep{yamauchi16} and updated luminosity distribution measurements of local DNe  \citep{byckling10,reis13}.

{\it Summary}. {\it NuSTAR} provides possibility to study emission from the Galactic center and bulge with the same instrument using both mirror's response and side aperture. The broad-band continuum of the bulge measured by \cite{perez19} using  {\it NuSTAR}'s side aperture is consistent with a dominant population of DNe, confirming the  detailed {\it Suzaku} studies of the Fe line properties and low-energy continuum of the bulge and updated luminosity distribution measurements of local DNe.

\acknowledgements 
E.K. thanks for support RFBR, project number 19-32-90283. R.K. acknowledges support from the Russian Science Foundation (grant 19-12-00396). K.P. receives support from the Alfred P. Sloan Foundation.

\bibliography{7c_kuznetsova}  

\end{document}